\documentstyle[12pt]{article}

\newcommand{\EQ}{\begin{equation}}
\newcommand{\EN}{\end{equation}}

\newcommand{\ket}[1]{\left|#1\right\rangle}      % Ket-Zustand
\newcommand{\bear}{\begin{eqnarray}}
\newcommand{\ear}{\end{eqnarray}}
\newcommand{\bt} { \begin{tabular} }
\newcommand{\et}{ \end{tabular} }
\newcommand{\bc} { \begin{center} }
\newcommand{\ec}{ \end{center} }

\newcommand{\btb} { \begin{table} }
\newcommand{\etb}{ \end{table} }
\begin{document}

\topmargin 0pt
\oddsidemargin 5mm
\newcommand{\NP}[1]{Nucl.\ Phys.\ {\bf #1}}
\newcommand{\PL}[1]{Phys.\ Lett.\ {\bf #1}}
\newcommand{\NC}[1]{Nuovo Cimento {\bf #1}}
\newcommand{\CMP}[1]{Comm.\ Math.\ Phys.\ {\bf #1}}
\newcommand{\PR}[1]{Phys.\ Rev.\ {\bf #1}}
\newcommand{\PRL}[1]{Phys.\ Rev.\ Lett.\ {\bf #1}}
\newcommand{\MPL}[1]{Mod.\ Phys.\ Lett.\ {\bf #1}}
\newcommand{\JETP}[1]{Sov.\ Phys.\ JETP {\bf #1}}
\newcommand{\TMP}[1]{Teor.\ Mat.\ Fiz.\ {\bf #1}}
     
\renewcommand{\thefootnote}{\fnsymbol{footnote}}
     
\newpage
\setcounter{page}{0}
\begin{titlepage}     
\begin{flushright}
UFSCARF-TH-98-12
\end{flushright}
\vspace{0.5cm}
\begin{center}
{\large Integrable mixed vertex models from braid-monoid algebra}\\
\vspace{1cm}
\vspace{1cm}
{\large $M.J.Martins$ } \\
\vspace{1cm}
{\em Universidade Federal de S\~ao Carlos \\
Departamento de F\'isica \\
C.P. 676, 13560~~S\~ao Carlos, Brasil}\\
\end{center}
\vspace{0.5cm}
     
\begin{abstract}
We use the braid-monoid algebra to construct integrable mixed
vertex models. The transfer matrix of a mixed $SU(N)$ model 
is diagonalized by nested Bethe ansatz approach.

\end{abstract}
\vspace{.15cm}
%\centerline{PACS numbers: 05.20-y, 05.50+q, 04.20.Jb, 03.65.Fd}
\vspace{.1cm}
\centerline{July 1998}
\end{titlepage}

\renewcommand{\thefootnote}{\arabic{footnote}}
\section{Introduction}

The quantum version of the inverse scattering method paved the way for the
discovery and solution of new exactly solvable two-dimensional
lattice models. Of particular interest are non-homogeneous vertex 
models whose transfer matrix are constructed by mixing different local
vertex operators. These operators, 
known as Lax  ${\cal L}$-operators, define
the local structure of Boltzmann weights of the
system. A sufficient condition for integrability is that all
the ${\cal L}$-operators should satisfy the Yang-Baxter equation with the
same invertible $R$-matrix. More precisely we have \cite{FA},
\EQ
R(\lambda - \mu) {\cal L}_{{\cal A} i}(\lambda) \otimes
{\cal L}_{{\cal A} i}(\mu) =
{\cal L}_{{\cal A} i}(\mu) \otimes
{\cal L}_{{\cal A} i}(\lambda) 
 R(\lambda - \mu)
\EN

The auxiliary space ${\cal A}$ corresponds to the horizontal degrees of
freedom of the vertex model in the square lattice. The operator
${\cal L}_{{\cal A} i}(\lambda)$  is a matrix in the auxiliary space ${\cal A}$
and its matrix elements are operators on the 
quantum space $ \displaystyle \prod_{i=1}^{L}
\otimes V_i$, where $V_i$ represents the vertical space of states  and $i$
the sites of 
a one-dimensional lattice of size $L$. The tensor
product in formula (1) is taken with respect the auxiliary space and $\lambda$
is a spectral parameter.
One way of producing a mixed vertex model is by choosing
${\cal L}$-operators intertwining between different 
representations $V_i$ of a given underlying algebra. This approach has first
been used by Andrei and Johannesson \cite{AN} to study the
Heisenberg model in the presence of an impurity of spin-S and by De Vega
and Woynarovich \cite{DW} to construct alternating Heisenberg
spin chains. Subsequently, several papers have discussed physical
properties of the latter models \cite{SH,MA,JA} as well as considered
generalizations to include other Lie algebras \cite{MA1,AB} and
superalgebras \cite{BE,AB1,LI,SH1}.

In this paper, which we are pleased to dedicate to
James McGuire, we construct and solve mixed
vertex models whose ${\cal L}$-operators can be expressed in terms of
the generators of the braid-monoid algebra \cite{WA}. In particular, we
show that the recent results by Abad and Rios \cite{AB,AB1} 
and by Links and Foerster \cite{LI} for the $SU(3)$ and $gl(2|1)$
algebras can be reobtained from such algebraic approach.
The novel
feature as compared to these works is that we are able to diagonalize
the corresponding transfer matrix by using a standard variant of the
nested Bethe ansatz approach. Furthermore, this algebraic approach
allows us to 
derive extensions to the $SU(N)$ and $Sl(N|M)$ algebras in a more
direct way.

This paper is organized as follows. In section 2 we recall the basics
of the braid-monoid algebra and show how it produces two
different ${\cal L}$-operators
satisfying the Yang-Baxter equation (1). In section 3
a mixed vertex model based on the $SU(N)$  algebra is
diagonalized by the algebraic Bethe ansatz method.

\section{Braid-monoid {\cal L}-operators}

It is well known that the braid algebra produces the simplest 
rational $R$-matrix
solution of the Yang-Baxter equation. In this case the braid operator becomes
a generator of the symmetric group and the $R$-matrix is given by
\EQ
R(\lambda) = I_i + \lambda b_i 
\EN
where $I_i$ is the identity and $b_i$ is the braid operator acting
on the sites $i$ and $i+1$ of a one-dimensional chain. Here we choose
the braid operator as the graded permutation between $N$ bosonic and
$M$ fermionic degrees of freedom \cite{KU}
\EQ
b_i  = \sum_{a,b=1}^{N+M}(-1)^{p(a)p(b)} e_{ab}^{i} \otimes e_{ba}^{i+1}
\EN
where $p(a)$ is the Grassmann parity of the $a$-th degree of freedom, 
assuming values $p(a)=0$  for bosons and $p(a)=1$ for fermions. 

The $R$-matrix (2,3) has a null Grassmann parity, and consequently can
produce a vertex operator
${\cal L}_{{\cal A} i}(\lambda) $ satisfying either the Yang-Baxter
equation or its graded version \cite{KU}. In the latter case, the tensor
product in formula (1) should be taken in the graded sense(supertensor
product) \cite{KU}. In the latter case the associated ${\cal L}$-operator is
\EQ
{\cal L}_{{\cal A} i}(\lambda) = \lambda I_i +b_i
\EN

The next step is to search for extra ${\cal L}$-operators which
should satisfy equation (1) with the $R$-matrix (2,3). As we shall see
below, this is possible when we enlarge the braid algebra by including
a Temperley-Lieb operator $E_i$. This operator satisfies the following
relations
\EQ
\begin{array}{l}
E_{i}^{2} = q E_{i},~~ 
E_{i}E_{i\pm1}E_{i} = E_{i},~~ 
E_{i}E_{j} = E_{j}E_{i}  \hspace{2cm} |i-j| \geq 2 \\
\end{array}
\EN
where $q$  is a $c$-number. It turns out that the braid operator (3) 
together with the monoid $E_i$ close the braid-monoid algebra \cite{WA} at
its degenerated point ($b_i^2=I_i$). The extra relations between $b_i$
and $E_i$ closing
the degenerated braid-monoid algebra are (see e.g \cite{PM})
\EQ
\begin{array}{l}
b_{i}E_{i} = E_{i}b_i= \hat{t} E_{i} \\
E_{i}b_{i\pm1}b_{i} = b_{i\pm1}b_{i}E_{i\pm1} = E_{i}E_{i\pm1} \\
\end{array}
\EN
where the constant $\hat{t}$ assumes the values $\pm1$. 

Now we have to solve the Yang-Baxter equation (1) with the $R$-matrix (2,3)
assuming the following general ansatz for the ${\cal L}$-operator
\EQ
{\cal L}_{{\cal A} i}(\lambda) = f(\lambda) I_i +g(\lambda)b_i +h(\lambda)E_i
\EN
where $f(\lambda)$, $g(\lambda)$ and $h(\lambda)$ are functions to
be determined as follows. Substituting this ansatz in equation (1), and
taking into account the braid-monoid relations, we find two classes
of solutions. The first one has $h(\lambda)=0$ and clearly corresponds
to the standard solution already given in equation (4). The second one is
new and is giving by $g(\lambda)=0$. We find that the new 
${\cal L}$-operator,
after normalizing the solution by function $h(\lambda)$, is
given by
\EQ
{\tilde{\cal L}}_{{\cal A} i}(\lambda) = \hat{t}(\lambda +\eta)I_i -E_i
\EN
where $\eta$ is an arbitrary constant. This constant
can be fixed  imposing unitary property, i.e 
${\tilde{\cal L}}_{{\cal A} i}(\lambda) 
{\tilde{\cal L}}_{{\cal A} i}(-\lambda) \sim I_i$. By using this property
and the first equation (5) we find
\EQ
\eta =\frac{q}{2 \hat{t}}
\EN

After having found two distincts ${\cal L}$-operators which satisfy the
Yang-Baxter algebra with the same $R$-matrix, the construction of
an integrable mixed vertex model becomes standard \cite{AN,DW}. The monodromy
matrix of a vertex model mixing $L_1$ operators of type 
${\cal L}_{{\cal A} i}(\lambda)$ and $L_2$ operators of type
${\tilde{\cal L}}_{{\cal A} i}(\lambda) $ is written as
\EQ
{\cal T}^{L_1,L_2}(\lambda) = {\bar{\cal L}}_{{\cal A} L_1+L_2}(\lambda)
{\bar{\cal L}}_{{\cal A} L_1+L_2-1}(\lambda) \cdots 
{\bar{\cal L}}_{{\cal A}1}(\lambda)
\EN
where 
${\bar{\cal L}}_{{\cal A}i}(\lambda)$ is defined by
\EQ
{\bar{\cal L}}_{{\cal A}i}(\lambda) = \left\{ \begin{array}{ll}
{\cal L}_{{\cal A}i}(\lambda) & \mbox{if $i \in \{ \beta_1, \cdots, 
\beta_{L_1} \} $ } \\
{\tilde{\cal L}}_{{\cal A}i}(\lambda) & \mbox{ otherwise} 
\end{array} \right.
\EN   
and the partition $ \{\beta_1, \cdots, \beta_{L_1} \} $ denotes 
a set of integer indices assuming values in the interval 
$1 \leq \alpha_i \leq L_1+L_2$. Although the integrability does not
depend on how we choose such partition, the construction of
$local$ conserved charges commuting with the
respective transfer matrix does. One interesting case is when the number
of operators
${\cal L}_{{\cal A}i}(\lambda)$ and
${\tilde{\cal L}}_{{\cal A}i}(\lambda)$ are equally distributed ($L_1=L_2=L$)
in an alternating way in the monodromy matrix \cite{DW}. In this case, the
first non-trivial charge, known as Hamiltonian, is given in terms
of nearest neighbor and next-to-nearest neighbor interactions. More 
specifically, the expression for the Hamiltonian in the absence of
fermionic degrees of freedom($M=0$) is
\EQ
H=  \sum_{m \in odd}^{2L} 
{\tilde{\cal L}}_{m-1,m}(0) +
\sum_{i \in even}^{2L} 
{\tilde{\cal L}}_{n-2,n-1}(0) 
{\tilde{\cal L}}_{n,n-1}(0) 
P_{n-2,n}
\EN
where in the computations it was essential to use the 
unitary property 
${\tilde{\cal L}}_{{\cal A}i}(0)
{\tilde{\cal L}}_{{\cal A}i}(0)= \frac{q^2}{4} I_i$ at the regular
point $\lambda=0$. We also recall that $P_{ij}$ denotes  permutation
between sites $i$ and $j$(equation (3) with $M=0$).

We close this section discussing explicit representations for the
monoid $E_i$. Such representations can be found in terms of the
invariants of the superalgebra $Osp(N|2M)$, where $N$ and $2M$ are
the number of bosonic and fermionic degrees of freedom, respectively.
Here we shall consider a representation that respects the $U(1)$
invariance, which will be very useful in Bethe ansatz analysis.
Following ref. \cite{PM} the monoid is written as
\EQ
E_{i} = \sum_{a,b,c,d=1}^{N+2M} \alpha_{ab} \alpha^{-1}_{cd}
e_{ac}^{i} \otimes e_{bd}^{i+1}
\EN
and the matrix $\alpha$ has the following block anti-diagonal
structure
\EQ
\alpha=\left( \begin{array}{ccc} 
	O_{N \times M
} &   O_{N \times M} & {\cal{I}}_{N \times N} \\
	O_{M \times M} &   {\cal{I}}_{M \times M
} & O_{M \times N} \\
	-{\cal{I}}_{M \times M} &   O_{M \times M} & O_{M \times N} \\
	\end{array}
	\right)
\EN
where 
${\cal{I}}_{ k_1 \times k_2} $ 
and 
${O}_{ k_1 \times k_2} $ are the anti-diagonal and the null  
 $ k_1 \times k_2 $ 
matrices, respectively. We also recall that for $\hat{t}=1$ the
sequence of grading is  
$ f_1 \cdots
f_M b_1 \cdots b_N f_{M+1} \cdots f_{2M}$ and the Temperley-Lieb
parameter $q$ is the difference between the number of bosonic and fermionic
degrees of freedom
\EQ
q=N-2M
\EN

Finally, we remark that new  
${\tilde{\cal L}}_{{\cal A}i}(\lambda)$ operators are obtained only
when $ N+2M \geq 3$. Indeed, for the special cases $N=2$, $M=0$ and
$N=0$, $M=1$ it is possible to verify that such operator has the structure
of the 6-vertex model, which is precisely the same of 
${\cal L}_{{\cal A}i}(\lambda)$, modulo trivial phases and scaling.
In the cases $N=3$,$M=0$ and $N=1$,$M=1$ we reproduce, after
a canonical transformation, the $\cal{L}$-operators used recently 
in the literature \cite{AB,AB1,LI} to construct mixed $SU(3)$ and
$t-J$ models. As we shall see in next
section, however, an important advantage of our approach
is that representation (14) is the appropriate one 
to allows us to perform standard
nested Bethe Ansatz diagonalization. 

\section{Bethe ansatz diagonalization}

In this section we look at the problem of diagonalization of
the transfer matrix 
$T^{L_1,L_2}(\lambda) =Tr_{{\cal A}}[ {\cal T}^{L_1,L_2}(\lambda)]$,
namely
\EQ
T^{L_1,L_2}(\lambda) \ket{\Phi} = \Lambda (\lambda) \ket{\Phi}
\EN
by means of the quantum inverse scattering method.

For sake of simplicity we  restrict ourselves to the case of mixed 
vertex models in the absence of fermionic degrees of freedom. In this
case the ${\cal L}$-operators 
${\cal L}_{{\cal A}i}(\lambda)$ and ${\tilde{\cal L}}_{{\cal A}i}(\lambda)$
are given by formulae (4) and (8) with $M=0$, $\hat{t}=1$ and $\eta=N/2$.
An important object in this framework is the reference state $\ket{0}$
we should start with in order to construct the full Hilbert space
$\ket{\Phi}$. The structure of the ${\cal L}$-operators suggests us
to take the standard ferromagnetic pseudovacuum as our reference state,
i.e
\EQ
\ket{0} = \prod_{i=1}^{L_1+L_2} \otimes \ket{0}_{i} , ~~
\ket{0}_{i} = 
\pmatrix{
1 \cr
0 \cr
\vdots \cr
0 \cr}_{N}
\EN
where the index $N$ represents the length of the vectors $\ket{0}_{i}$. 
It turns out that this state is an exact eigenvector of the 
transfer matrix,
since both operators
${\cal L}_{{\cal A}i}(\lambda)$ and ${\tilde{\cal L}}_{{\cal A}i}(\lambda)$
satisfy the following important triangular properties
\EQ 
{\cal L}_{{\cal A}i}(\lambda)\ket{0}_{i} = 
\pmatrix{
a(\lambda) \ket{0}_i &  *  &  *   & \dots & *   \cr
0  &  b(\lambda) \ket{0}_i  &  0  & \dots & 0  \cr
\vdots & \vdots & \ddots & \dots & \vdots  \cr
0  &  0  &  0  & \dots & b(\lambda)\ket{0}_i  \cr}_{N \times N}
\EN
and 
\EQ 
{\tilde{\cal L}}_{{\cal A}i}(\lambda)\ket{0}_{i} = 
\pmatrix{
{\tilde b}(\lambda) \ket{0}_i &  0  &  0   & \dots & *   \cr
0  &  {\tilde b}(\lambda) \ket{0}_i  &  0  & \dots & *  \cr
\vdots & \vdots & \ddots & \dots & \vdots  \cr
0  &  0  &  0  & \dots & {\tilde a}(\lambda)\ket{0}_i  \cr}_{N \times N}
\EN
where the symbol $*$ stands for non-null values  that are not necessary to 
evaluate in this algebraic approach. The functions $a(\lambda)$, $b(\lambda)$,
$\tilde{a}(\lambda)$ and $\tilde{b}(\lambda)$ are obtained directly
from expressions (8,9), and they are given by
\EQ
a(\lambda)=\lambda+1,~~b(\lambda)=\lambda,~~\tilde{a}(\lambda)=
\lambda+\frac{N}{2} -1,~~\tilde{b}(\lambda)=\lambda+ \frac{N}{2}
\EN

To make further progress we have to write an appropriate ansatz for the
monodromy matrix
${\cal T}^{L_1,L_2}(\lambda)$ in the auxiliary space ${\cal A}$. The
triangular properties (18,19) suggest us to seek for  standard structure
used in nested Bethe ansatz 
diagonalization of $SU(N)$ vertex models \cite{DV},
\EQ
{\cal T}^{L_1,L_2}(\lambda) = 
\pmatrix{
A(\lambda)       &   B_i(\lambda)   \cr
C_i(\lambda)  &  D_{ij}(\lambda)  \cr}_{N \times N}
\EN
where $i,j=1, \cdots, N-1$. As a consequence of properties (18,19) we 
derive how the monodromy matrix elements act on the reference state. 
The fields $B_i(\lambda)$ play the role of creation operators
while $C_i(\lambda)$ are annihilation fields, i.e 
$C_i(\lambda) \ket{0}=0$. Furthermore, the action of the ``diagonal''
operators $A(\lambda)$ and $D_{ij}(\lambda)$ are given by
\EQ
A(\lambda)\ket{0} = [a(\lambda)]^{L_1}
[\tilde{b}(\lambda)]^{L_2}\ket{0} 
\EN
\EQ 
D_{ii}(\lambda)\ket{0} = 
[b(\lambda)]^{L_1}
[\tilde{b}(\lambda)]^{L_2}\ket{0} \mbox{ for $i \neq N-1$},~~
D_{N-1,N-1}(\lambda)\ket{0} = 
[b(\lambda)]^{L_1}
[\tilde{a}(\lambda)]^{L_2}\ket{0} 
\EN
\EQ 
D_{i,j}(\lambda)\ket{0} =0  \mbox{ for $i \neq j$ and $j \neq N-1$},~~ 
D_{i,N-1}(\lambda)\ket{0}  \neq 0 \mbox{ for $i \neq N-1$}
\EN

We observe that although matrix 
$D_{ij}(\lambda) \ket{0}$ is non-diagonal it is
up triangular, which is an important property to carry on higher level
Bethe ansatz analysis. In order to construct other eigenvectors we
need to use the commutation relations between the monodromy matrix elements
which are obtained by extending the Yang-Baxter relation to the monodromy
matrix ansatz (21). Due to the structure of the $R$-matrix, the commutation
rules are the same of that already known for isotropic $SU(N)$ models \cite{DV},
and the most useful relations for subsequent derivations are
\EQ
A(\lambda) B_i(\mu) = 
\frac{a(\mu-\lambda)}{b(\mu-\lambda)}B_i(\mu) A(\lambda) 
- \frac{1}{b(\mu-\lambda)} B_i(\lambda) A(\mu)   
\EN
\EQ
D_{ij}(\lambda)B_k(\mu) = 
\frac{1}{b(\lambda-\mu)} B_{p}(\mu)D_{iq}(\lambda) r^{(1)}(\lambda-\mu)_{pq}^{jk} 
-\frac{1}{b(\lambda-\mu)} B_j(\lambda)D_{ik}(\mu)
\EN
\EQ
B_i(\lambda)B_j(\mu) = 
B_p(\mu)B_q(\lambda) r^{(1)}(\lambda-\mu)_{pq}^{ij} 
\EN
where $r^{(1)}(\lambda)_{pq}^{ij}$  are the elements of the $R$-matrix 
$I_i +\lambda b_i$ on the subspace $(N-1) \times (N-1)$. The eigenvectors
are given in terms of the following linear combination \cite{DV} 
\EQ
\ket{\Phi_{m_{1}}(\lambda^{(1)}_{1}, \cdots , \lambda^{(1)}_{m_{1}})} =  
B_{a_1}(\lambda^{(1)}_{1}) \cdots B_{a_{m_1}}(\lambda^{(1)}_{m_1})
{\cal F}^{a_{m_{1}} \cdots a_{1}}
\EN
where  the components 
${\cal F}^{a_{m_{1}} \cdots a_{1}}$ are going to be determined a posteriori.

By carring on the diagonal fields  $A(\lambda)$ and $D_{ii}(\lambda)$
over the above $m_1$-particle state
we generate   the so-called wanted and unwanted terms. The wanted terms
are those proportional to
$\ket{\Phi_{m_{1}}(\lambda^{(1)}_{1}, \cdots , \lambda^{(1)}_{m_{1}})}$ and
they contribute directly to the eigenvalue 
$\Lambda^{L_1,L_2}(\lambda,\{\lambda^{(1)}_{i}\})$. 
These terms are
easily obtained by keeping only the first term of the commutation 
rules (25,26) each time we turn $A(\lambda)$ and $D_{ii}(\lambda)$ over
one of the $B_{a_i}(\lambda^{(1)}_{i})$ component. The result of this
computations leads us to the following expression
\bear
T^{L_1,L_2}(\lambda) 
\ket{\Phi_{m_{1}}(\lambda^{(1)}_{1}, \cdots , \lambda^{(1)}_{m_{1}})} &=&
[a(\lambda)]^{L_1}
[\tilde{b}(\lambda)]^{L_2} \prod_{i=1}^{m_1} 
\frac{a(\lambda^{(1)}_i -\lambda)}{b(\lambda^{(1)}_i -\lambda)}
\ket{\Phi_{m_{1}}(\lambda^{(1)}_{1}, \cdots , \lambda^{(1)}_{m_{1}})}
\nonumber\\
&&+\prod_{i=1}^{m_1} 
\frac{1}{b(\lambda- \lambda^{(1)}_i)}
B_{b_{1}}(\lambda^{(1)}_{1})
\cdots B_{b_{m_{1}}}( \lambda^{(1)}_{m_{1}}) 
T^{(1)}(\lambda,\{\lambda^{(1)}_{j}\})_{b_{1} \cdots b_{m_{1}}}^{a_{1} 
\cdots a_{m_{1}}} {\cal F}^{a_{m_{1}} \cdots a_{1}} 
\nonumber\\
&&+ \mathrm{ unwanted~ terms}
\ear
where 
$T^{(1)}(\lambda,\{\lambda^{(1)}_{j}\}) $ is the transfer matrix
of the following inhomogeneous auxiliary vertex model
\EQ
T^{(1)}(\lambda,\{\lambda^{(1)}_{i}\})_{b_{1} \cdots b_{m_{1}}}^{a_{1} \cdots a_{m_{1}}} = 
r^{(1)}(\lambda-\lambda^{(1)}_1)_{b_{1}d_{1}}^{aa_{1}}
r^{(1)}(\lambda-\lambda^{(1)}_2)_{b_{2}d_{2}}^{d_{1}a_{2}} \cdots
r^{(1)}(\lambda-\lambda^{(1)}_{m_1})_{b_{m_{1}
}d_{m_1}}^{d_{m_1-1}a_{m_{1}}} D_{ad_{m_1}}(\lambda) \ket{0}
\EN

The unwanted terms arise when one of the variables 
$\lambda^{(1)}_i$ of the $m_1$-particle state 
is exchanged with the spectral parameter $\lambda$. It is known \cite{DV}
how to collect these in a close form, thanks to the commutation rule (27)
which makes possible to relate different ordered multiparticle states. 
We find that the unwanted terms of kind 
$B_{a_{1}}(\lambda^{(1)}_{1})
\cdots B_{a_{i}}(\lambda) \cdots B_{a_{m_{1}}}( \lambda^{(1)}_{m_{1}}) $
are cancelled out provided  we impose further restriction to the
$m_1$-particle state rapidities $\lambda_{i}^{(1)}$, namely
\bear
&&\left[a(\lambda^{(1)}_{i}) \right]^{L_1} 
\left[\tilde{b}(\lambda^{(1)}_{i}) \right]^{L_2}
\prod_{j=1 \; j \neq i}^{m_{1}} b(\lambda^{(1)}_{i}-\lambda^{(1)}_{j})
\frac{a(\lambda^{(1)}_{j}-\lambda^{(1)}_{i})}{b(\lambda^{(1)}_{j}-\lambda^{(1)}_{i})}
{\cal F}^{a_{m_{1}} \dots a_{1}}=
\nonumber \\
&& T^{(1)}(\lambda=\lambda^{(1)}_{i},\{\lambda^{(1)}_{j}\})^{b_{1} \cdots b_{m_{1}}}_{a_{1} \cdots a_{m_{1}}} {\cal F}^{b_{m_{1}} \dots b_{1}} , i=1, \dots, m_{1}
\ear

Now it becomes necessary to introduce a second Bethe ansatz in
order to diagonalize the auxiliary transfer matrix 
$T^{(1)}(\lambda,\{\lambda^{(1)}_{i}\})$. The only difference as compare
to standard cases \cite{DV} is the presence of the ``gauge'' matrix
$g_{ab} =D_{ab}(\lambda) \ket{0}$. It turns out that this problem is
still integrable since the tensor product $g \otimes g$ commutes
with the auxiliary $R$-matrix 
$r^{(1)}(\lambda)$ \footnote{ This occurs because
the off-diagonal elements $D_{i,N-1}(\lambda)$ belongs to
a commutative ring.}(see e.g ref. \cite{DV1}). From 
equations (23,24) we also note that this gauge does not spoil the
triangular form of the monodromy matrix associated
to
$T^{(1)}(\lambda,\{\lambda^{(1)}_{i}\})$ when it acts on the
usual ferromagnetic state,
\EQ 
\ket{0^{(1)}} = \prod_{i=1}^{m_1} \otimes
\pmatrix{
1 \cr
0 \cr
\vdots \cr
0 \cr}_{N-1} 
\EN

By defining $\Lambda^{(1)}(\lambda,\{\lambda^{(1)}_{i}\})$ as the 
eigenvalue of the auxiliary transfer matrix 
$T^{(1)}(\lambda,\{\lambda^{(1)}_{i}\})$, i.e  
\EQ
T^{(1)}(\lambda,\{\lambda^{(1)}_{i}\})^{b_{1} \cdots b_{m_{1}}}_{a_{1} \cdots a_{m_{1}}} {\cal F}^{b_{m_{1}} \dots b_{1}} =
\Lambda^{(1)}(\lambda,\{\lambda^{(1)}_{i}\}) {\cal F}^{a_{m_{1}} \dots a_{1}}
\EN
we derive from equation (29) that the eigenvalue of $T^{L_1,L_2}(\lambda)$
is given by
\EQ
\Lambda^{L_1,L_2}(\lambda,\{\lambda^{(1)}_{i}\})  =
[a(\lambda)]^{L_1} [\tilde{b}(\lambda)]^{L_2}
\prod_{i=1}^{m_{1}} \frac{a(\lambda^{(1)}_{i}-\lambda)}
{b(\lambda^{(1)}_{i}-\lambda)} 
+ \prod_{i=1}^{m_{1}}
\frac{1}{b(\lambda-\lambda^{(1)}_{i})}
\Lambda^{(1)}(\lambda,\{\lambda^{(1)}_{i}\})  
\nonumber\\
\EN
and the nested Bethe ansatz equations (31) become
\EQ
\left[ a(\lambda^{(1)}_{i}) \right]^{L_1} 
\left[ \tilde{b}(\lambda^{(1)}_{i}) \right]^{L_2}
\prod_{j=1 \; j \neq i}^{m_{1}} b(\lambda^{(1)}_{i}-\lambda^{(1)}_{j})
\frac{a(\lambda^{(1)}_{j}-\lambda^{(1)}_{i})}{b(\lambda^{(1)}_{j}-\lambda^{(1)}_{i})} =
\Lambda^{(1)}(\lambda = \lambda^{(1)}_{i},\{\lambda^{(1)}_{j}\})
,~~ i=1, \dots, m_{1}
\EN

In order to find the auxiliary eigenvalue 
$\Lambda^{(1)}(\lambda = \lambda^{(1)}_{i},\{\lambda^{(1)}_{j}\})$ we have
to introduce a new set of variables 
$\{ \lambda_1^{(2)}, \cdots, \lambda_{m_2}^{(2)} \}$ which parametrize the
eigenvectors of 
$T^{(1)}(\lambda,\{\lambda^{(1)}_{i}\})$. The structure of the
commutations rules as well as the eigenvector ansatz (28) remains basically
the same, and the expression for 
$\Lambda^{(1)}(\lambda = \lambda^{(1)}_{i},\{\lambda^{(1)}_{j}\})$ will
again depend on another auxiliary inhomogeneous vertex model having $(N-2)$
states per link. We repeat this procedure until we reach the 
$(N-2)$th step, where the auxiliary problem becomes of $6$-vertex type.
Since this nesting approach is well known in the literature \cite{DV},
we here only present our final results. The
eigenvalue of the transfer matrix $T^{L_1,L_2}(\lambda)$ is given by
\bear
\Lambda^{L_1,L_2} \left(\lambda;\{\lambda^{(1)}_j\}, 
\cdots,\{\lambda^{(N-1)}_j\} \right)& = &
[a(\lambda)]^{L_1}
[\tilde{b}(\lambda)]^{L_2}
\prod_{j=1}^{m_{1}} \frac{a(\lambda^{(1)}_{j}-\lambda)}
{b(\lambda^{(1)}_{j}-\lambda)}
\nonumber\\
&&+[b(\lambda)]^{L_1}
[\tilde{b}(\lambda)]^{L_2}
\sum_{l=1}^{N-2} 
\prod_{j=1}^{m_{l}} \frac{a(\lambda-\lambda^{(l)}_{j})}
{b(\lambda- \lambda^{(l)}_{j})}
\prod_{j=1}^{m_{l+1}} \frac{a(\lambda^{(l+1)}_{j}-\lambda)}
{b(\lambda^{(l+1)}_{j}-\lambda)}
\nonumber\\
&&+[b(\lambda)]^{L_1}
[\tilde{a}(\lambda)]^{L_2}
\prod_{j=1}^{m_{N-1}} \frac{a(\lambda-\lambda^{(N-1)}_{j})}
{b(\lambda-\lambda^{(N-1)}_{j})}
\ear
while the nested Bethe ansatz equations are given by
\bear
\left[ \frac{a(\lambda^{(1)}_{i})}
{b(\lambda^{(1)}_{i})} \right]^{L_1} =
\prod_{j=1,j \neq i}^{m_{1}} -\frac{a(\lambda^{(1)}_{i}-\lambda^{(1)}_{j})}
{a(\lambda^{(1)}_{j}-\lambda^{(1)}_{i})}
\prod_{j=1}^{m_{2}} \frac{a(\lambda^{(2)}_{j}-\lambda^{(1)}_{i})}
{b(\lambda^{(2)}_{j}-\lambda^{(1)}_{i})}
\nonumber\\
\ear
\bear
\prod_{j=1,j \neq i}^{m_{l}} -\frac{a(\lambda^{(l)}_{i}-\lambda^{(l)}_{j})}
{a(\lambda^{(l)}_{j}-\lambda^{(l)}_{i})} =
\prod_{j=1}^{m_{l-1}} \frac{a(\lambda^{(l)}_{i}-\lambda^{(l-1)}_{j})}
{b(\lambda^{(l)}_{i}-\lambda^{(l-1)}_{j})} 
\prod_{j=1}^{m_{l+1}} \frac{b(\lambda^{(l+1)}_{j}-\lambda^{(l)}_{i})}
{a(\lambda^{(l+1)}_{j}-\lambda^{(l)}_{i})},~ 
l= 2,\cdots, N-2
\nonumber\\
\ear
\bear
\left[ \frac{\tilde{a}(\lambda^{(N-1)}_{i})}
{\tilde{b}(\lambda^{(N-1)}_{i})} \right]^{L_2} =
\prod_{j=1,j \neq i}^{m_{N-1}} -\frac{a(\lambda^{(N-1)}_{j}-\lambda^{(N-1)}_{i})}
{a(\lambda^{(N-1)}_{i}-\lambda^{(N-1)}_{j})}
\prod_{j=1}^{m_{N-2}} \frac{a(\lambda^{(N-1)}_{i}-\lambda^{(N-2)}_{j})}
{b(\lambda^{(N-1)}_{i}-\lambda^{(N-2)}_{j})}
\ear

We would like  to close this paper with the following remarks. First we
note that it is possible to perform convenient shifts in the
Bethe ansatz rapidities, $\{ \lambda_i^{(p)} \} \rightarrow 
\{ \lambda_i^{(p)} \} -\frac{p}{2} $, in order to present the results
(36-39) in a more symmetrical form. For instance, after these shifts, 
the nested Bethe ansatz equations can be compactly written as
\EQ
\left[
\frac{\lambda^{(a)}_{i} -\frac{\delta_{a,w}}{2}}{\lambda^{(a)}_{i} +\frac{\delta_{a,w}}{2}} 
\right]^{L_w} =
\prod_{b=1}^{r} \prod_{k=1,\; k \neq i}^{m_{b}}
\frac{\lambda^{(a)}_{i}-\lambda^{(b)}_{k} -\frac{C_{a,b}}{2}}{\lambda^{(a)}_{i}-\lambda^{(b)}_{k} +\frac{C_{a,b}}{2}}, ~~ i=1, \dots, m_{a} ;~~ a=1, \dots,N-1 
\EN
where $C_{ab}$ is the Cartan matrix of the $A_N$
Lie algebra and $w=1,N-1$. We note that for $N=3$ we recover the results
by Abad and Rios \cite{AB}. It is also straightforward  to extend the above
Bethe ansatz 
results to the superalgebra $Sl(N|M)$. The $N=0$ case is the simplest
one, since the only modification is the addition of minus signs in functions
$\tilde{a}(\lambda)$ and $\tilde{b}(\lambda)$.

Next remark concerns  the physical meaning of the ${\cal L}$-operators
as  scattering $S$-matrices. It is known that 
${\cal L}_{{\cal A}i}(\lambda)$ 
might represent the 
scattering matrix of particles belonging to the
fundamental representation of $SU(N)$. The extra solution
$\tilde{{\cal L}}_{{\cal A}i}(\lambda)$, however, 
should be seen  as the forward scattering amplitude between a particle and
an antiparticle. In fact, it is possible to show that the whole 
particle-antiparticle scattering(even backward amplitudes)
can be closed in terms of the
braid-monoid algebra. We leave a detailed analysis of this
possibility, their Bethe ansatz properties as well as  generalizations
to include trigonometric solutions for a forthcoming paper \cite{MA2}.

\section*{Acknowledgements}
This work was supported by   
Fapesp ( Funda\c c\~ao
de Amparo \`a Pesquisa do Estado de S. Paulo) and Cnpq (Conselho Nacional
de Desenvolvimento Cient\'ifico e Tecnol\'ogico).

\end{document}